\documentclass[iop]{emulateapj}

\newcounter{species}
\def\ion#1#2{\setcounter{species}{#2}#1$\;${\scriptsize\Roman{species}}\relax}

\usepackage{hyperref}
\hypersetup{colorlinks,linkcolor=blue, citecolor=blue}

\usepackage{url}
\usepackage{color}
\shorttitle{Metallicity of CM Dra}
\shortauthors{Ryan C. Terrien et al.}
\begin{document}
\title{The Metallicity of the CM Draconis System}

\author{Ryan C. Terrien\altaffilmark{1,2},
Scott W. Fleming\altaffilmark{1,2},
Suvrath Mahadevan\altaffilmark{1,2},
Rohit Deshpande\altaffilmark{1,2},
Gregory A. Feiden\altaffilmark{3,4},
Chad F. Bender\altaffilmark{1,2},
Lawrence W. Ramsey\altaffilmark{1,2}}

\email{rct151@psu.edu}
\altaffiltext{1}{Department of Astronomy and Astrophysics, The Pennsylvania State University, 525 Davey Laboratory, University Park, PA 16802, USA.}
\altaffiltext{2}{Center for Exoplanets and Habitable Worlds, The Pennsylvania State University, University Park, PA 16802, USA.}
\altaffiltext{3}{Neukom Graduate Fellow}
\altaffiltext{4}{Department of Physics and Astronomy, Dartmouth College, 6127 Wilder Laboratory, Hanover, NH 03755}

\begin{abstract}
The CM Draconis system comprises two eclipsing mid-M dwarfs of nearly equal mass in a 1.27-day orbit.  This well-studied eclipsing binary has often been used for benchmark tests of stellar models, since its components are amongst the lowest mass stars with well-measured masses and radii ($\lesssim 1$\% relative precision). However, as with many other low-mass stars, non-magnetic models have been unable to match the observed radii and effective temperatures for CM Dra at the 5-10\% level. To date, the uncertain metallicity of the system has complicated comparison of theoretical isochrones with observations.  In this Letter, we use data from the SpeX instrument on the NASA Infrared Telescope Facility (IRTF) to measure the metallicity of the system during primary and secondary eclipses, as well as out of eclipse, based on an empirical metallicity calibration in the $H$ and $K$ near-infrared (NIR) bands. We derive a $\rm{[Fe/H]} = -0.30 \pm 0.12$ that is consistent across all orbital phases.  The determination of $\rm{[Fe/H]}$ for this system constrains a key dimension of parameter space when attempting to reconcile model isochrone predictions and observations.
\end{abstract}

\section{Introduction}
\label{introsection}
CM Draconis (GJ 630.1 AC, \citealt{gli1991}, hereafter CM Dra) is an eclipsing binary system that consists of a pair of very similar M dwarfs. The eclipses are nearly total (89\% and 99.9\% of light is blocked from the primary and secondary, respectively). The first spectroscopic mass measurements were made by \citet{lac1977}, who noted that the two stars did not lie near the theoretical zero-age main sequence for Population I stars.  A more recent analysis by \citet{mor2009} has determined a primary mass $M_1 = 0.2310 \pm 0.0009 \, M_{\odot}$ and radius $R_1 = 0.2534 \pm 0.0019 \, R_{\odot}$, with a secondary mass $M_2 = 0.2141 \pm 0.0010 \, M_{\odot}$ and radius $R_2 = 0.2396 \pm 0.0015 \, R_{\odot}$.  The system's appreciable proper motion, relatively large radial velocity, and the existence of a common proper motion white dwarf companion have been claimed as evidence that the system may be a Population II member \citep{lac1977}.

Observationally determined radii and effective temperatures of low-mass stars ($M \lesssim 0.8 \: M_{\odot}$) disagree significantly with theoretical isochrones \citep{lop2007, mor2008, tor2010, sta2012}, and CM Dra is no exception \citep{mor2009, fei2011}. This disagreement was generally at the 5\% - 15\% level, but a more recent study by \citep{fei2012} uses a finely sampled grid of stellar parameters and an improved equation of state compared to previous studies, reducing the disagreement to the sub-5\% level for a majority of systems. Magnetic inhibition of convective efficiency or an increase of cool starspots at polar latitudes are thought to be the primary cause \citep{cha2007,mor2010,fei2012} of this discrepancy, but unknown parameters such as age and metallicity make it difficult to compare observations with theory. In the case of CM Dra, models attempting to mimic the interactions of a magnetic field with the stellar interior can be made to match the observations, but often require a super-solar metallicity, a large starspot coverage, or very strong magnetic fields \citep{mor2010,mac2012}.

The metallicity of the CM Dra system merits special consideration because it can constrain one dimension of the modeling problem for a system in the fully-convective regime ($\lesssim 0.35 \, M_{\odot}$) where there are only four other well-characterized systems: KOI-126 \citep{car2011}, Kepler 16 \citep{doy2011,ben2012}, Kepler 38 \citep{oro2012}, and LSPM J1112+7626 \citep{irw2011}. In the first three cases, metallicity estimates can be obtained from their higher-mass primary stars, but no such primary is available for CM Dra, and so metallicities must be determined from the low-mass stars alone.

Attempts to measure the metallicity of CM Dra using synthetic spectral templates in both the optical and near-infrared (NIR) have resulted in metallicity values ranging from solar \citep{giz1997} to $\rm{[M/H]} = -1.0$ \citep{vit2002}.  However, such techniques rely on model atmospheres of cool stars that have incomplete line lists and significant uncertainties in the complicated molecular chemistry occurring in the outer atmospheres.  An alternative approach is to use an empirical relationship based on equivalent widths of specific, narrow regions of the spectrum \citep{roj2010,roj2012,ter2012}.

In this Letter, we utilize such an empirical relation to derive the $\rm{[Fe/H]}$ of CM Dra from spectra taken at a variety of orbital phases:  primary eclipse, secondary eclipse, and out-of-eclipse.  By using a technique that does not rely heavily on low-mass stellar atmosphere models, and obtaining observations of the near-total eclipses that isolate the light from one star, we can obtain a reliable estimate of CM Dra's metallicity.  In Section \ref{obssection} we describe our SpeX data and the empirical relation used to derive metallicities.  In Section \ref{resultsection} we derive our metallicity estimate and compare it to values in the literature. In Section \ref{modelsection} we discuss the impact of this metallicity constraint on stellar models of the CM Dra system. Finally, in Section \ref{conclusionsection} we summarize our results.

\section{Observations with the IRTF-SpeX Spectrograph}
\label{obssection}
During May 2012, we observed CM Dra with the NASA Infrared Telescope Facility (IRTF) SpeX spectrograph \citep{ray2003} as a part of a larger program to estimate metallicities for several hundred M dwarfs. As part of this larger program, we have developed independent calibrations of metallicity ([Fe/H]) for the $K$ and $H$-bands of IRTF-SpeX M dwarf spectra \citep{ter2012}. We operated SpeX in the short cross-dispersed (SXD) mode with the $0.3 \times 5''$ slit, which produces $R \sim 2000$ spectra from 0.8 - 2.4 $\mu\rm{m}$. We extract these spectra with the facility SpeXTool package \citep{cus2004}. We telluric-correct and flux-calibrate the spectra using the {\it xtellcor} program \citep{vac2003} and observations of an A0V star (or similar), at an airmass and observation time within 0.1 ($\sec{z}$) and 1 hour of the target, respectively. Our analysis of the reduced spectra is identical to that presented in \citet{ter2012}, but here we note a few important points.

Our $K$ and $H$-band metallicity relations are based on the equivalent widths of \ion{Na}{1} and \ion{Ca}{1} in the $K$-band and \ion{K}{1} and \ion{Ca}{1} in the $H$-band. The windows used to calculate the strengths of these features are defined and explained in \citet{ter2012}. These windows are shifted for each observation by a radial velocity (RV) offset derived using a $K$-band spectral template \citep[a high S/N spectrum of HD 36395 from][]{ray2009}.\footnote{http://irtfweb.ifa.hawaii.edu/$\sim$spex/IRTF\_Spectral\_Library/} The measured equivalent widths are not sensitive to rotational broadening for a typical M dwarf rotational velocity ($\lesssim$ 30 km/s), as the rotational broadening is small compared to the spectrograph resolution (R $\sim$ 2000). For CM Dra in particular, \citet{mor2009} used visible light echelle spectra with a spectroscopic resolution of $\sim35000$ to measure $v \sin i \approx 10$ km/s for both components. We have simulated the impact of this $v \sin i$ on the \citet{ter2012} windows, and find it affects our measurements of equivalent width and metallicity at less than 0.1\%.

The metallicity relations are calibrated using a set of 22 M dwarf companions to higher-mass stars that have well-determined metallicities from the SPOCS catalog \citep{val2005}, and yield [Fe/H], which can be used as a proxy for [M/H]. We have demonstrated that the two independent calibrations are consistent with one another \citep{ter2012}. Since the features in the $H$-band are weaker than those used in the $K$-band, the $H$-band calibration requires S/N $\approx 150\: \rm{pixel}^{-1}$ while the $K$-band requires S/N $\approx 100\: \rm{pixel}^{-1}$. Our adopted error in the calibrations is $\pm 0.12$ dex, based on the the scatter of the metallicity estimates compared to the metallicity measurements of the primaries.

\begin{figure}
\begin{center}
\includegraphics[scale=0.45]{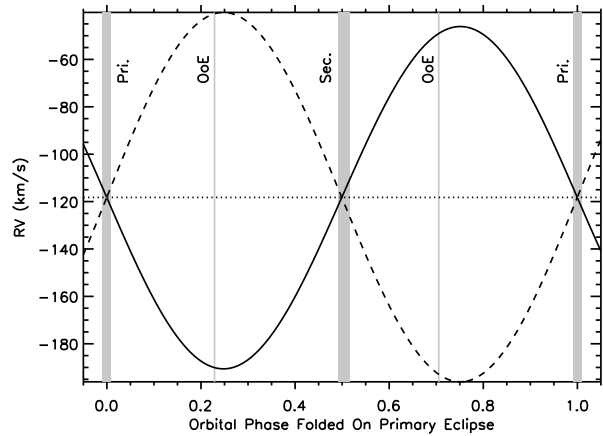}
\caption{RV model \citep[using RVLIN,][]{wri2009} of CM Dra using orbital parameters from \citet{mor2009}, showing the primary (solid line), secondary (dashed line), and gamma velocity (dotted line). The gray bars mark the phases at which our IRTF spectra were obtained: primary eclipse (Pri.), secondary eclipse (Sec.), and out-of-eclipse (OoE).\label{obsphases}}
\end{center}
\end{figure}

We observed CM Dra over a period of several days, distributing observations over many phases, including one primary and two secondary eclipses, as well as two out-of-eclipse observations. Both of our out-of-eclipse observations occurred close to maximum RV separation (Figure \ref{obsphases}), have high S/N ($> 300$), and contain full light from both components of CM Dra. We planned the three in-eclipse observations using the orbital parameters from \citet{mor2009}. We obtained coverage for 30-40 minutes, centered on the times of maximum eclipse, by taking repeated 30 second exposures and combining them into $\sim 2$ minute exposures during extraction. Both primary and secondary eclipses for CM Dra last a total of $\sim$ 1.5 hours. At the midpoint of the primary eclipse, the primary is $\sim 89$\% obscured; at the midpoint of the secondary eclipse, the secondary is $\sim 99.9$\% obscured. We can empirically estimate the effect of spectral contamination from either component on our metallicity measurements by observing during both eclipses and out of eclipse.

Around the time of maximum secondary eclipse, the secondary remains more than 98\% obscured for approximately three minutes, and we obtained the S/N $\geq 150$ required to perform both the $K$ and $H$-band metallicity estimates. Since our calibration is valid only for single stars, the isolated light from the primary allows us to estimate the metallicity of this component of the CM Dra system. Assuming the stars are coeval, this then provides our best metallicity estimate for the CM Dra system.

\section{Results and Discussion}
\label{resultsection}
Table \ref{restable} lists our individual CM Dra metallicity measurements and Table \ref{summarytable} summarizes them by epoch. The combined $K$ and $H$-band measurements for the secondary eclipse (effectively the metallicity estimate of the primary), as split into 2-minute exposures, have a mean of [Fe/H]$ = -0.30$ dex with a spread of $\sigma = 0.03$ dex, which is much smaller than the absolute uncertainty in our metallicity calibration relationship (Section \ref{obssection}). We therefore adopt the uncertainty in our calibration as the error for the metallicity estimate for CM Dra. We find CM Dra to be slightly metal-poor relative to the Sun, with the caveat that we are not sensitive to abundance patterns which could yield differences between [Fe/H] and [M/H].

\begin{deluxetable*}{ccccccccccc}
\tabletypesize{\small}
%\rotate{}
\tablecaption{CM Dra Metallicity Measurements\label{restable}}
\tablewidth{0pt}
\tablehead{
&&\multicolumn{4}{c}{$K$-band} & \multicolumn{4}{c}{$H$-band} \\
\colhead{RJD\tablenotemark{a}} & \colhead{In Eclipse?} & \colhead{Phase\tablenotemark{b}} & \colhead{EW$_{\mathrm{Na}}$[\AA]} & \colhead{EW$_{\mathrm{Ca}}$[\AA]} & \colhead{S/N$_{\mathrm{K}}$} & \colhead{[Fe/H]$_{\mathrm{K}}$} & \colhead{EW$_{\mathrm{Ca}}$[\AA]} &\colhead{EW$_{\mathrm{K}}$[\AA]} & \colhead{S/N$_{\mathrm{H}}$} & \colhead{[Fe/H]$_{\mathrm{H}}$}
}
\startdata
 56055.03708 & No & 0.229 &  3.58 $\pm$  0.07 &  2.28 $\pm$  0.12 &  312 & -0.30 &  1.25 $\pm$  0.11 &  0.57 $\pm$  0.05 &  384 & -0.40 \\
\hline
 56056.90978 & No & 0.706 &  3.44 $\pm$  0.06 &  2.20 $\pm$  0.07 &  344 & -0.33 &  1.38 $\pm$  0.04 &  0.63 $\pm$  0.02 &  404 & -0.37 \\
\hline
 56050.93011 & Primary & 0.992 &  3.45 $\pm$  0.08 &  2.27 $\pm$  0.10 &  246 & -0.32 &  1.06 $\pm$  0.07 &  0.78 $\pm$  0.04 &  274 & -0.38 \\
 56050.93380 & Primary & 0.994 &  3.46 $\pm$  0.07 &  2.25 $\pm$  0.09 &  260 & -0.32 &  1.06 $\pm$  0.07 &  0.81 $\pm$  0.03 &  285 & -0.36 \\
 56050.93749 & Primary & 0.997 &  3.42 $\pm$  0.08 &  2.23 $\pm$  0.10 &  243 & -0.33 &  1.12 $\pm$  0.06 &  0.84 $\pm$  0.03 &  264 & -0.34 \\
 56050.94118 & Primary & 0.000 &  3.40 $\pm$  0.07 &  1.97 $\pm$  0.10 &  235 & -0.35 &  0.95 $\pm$  0.06 &  0.77 $\pm$  0.04 &  257 & -0.38 \\
 56050.94486 & Primary & 0.003 &  3.16 $\pm$  0.08 &  2.23 $\pm$  0.08 &  245 & -0.36 &  0.97 $\pm$  0.08 &  0.79 $\pm$  0.03 &  268 & -0.36 \\
 56050.94855 & Primary & 0.006 &  3.39 $\pm$  0.07 &  2.31 $\pm$  0.08 &  277 & -0.32 &  1.03 $\pm$  0.05 &  0.78 $\pm$  0.03 &  305 & -0.37 \\
 56050.95316 & Primary & 0.010 &  3.53 $\pm$  0.06 &  2.21 $\pm$  0.06 &  368 & -0.31 &  1.03 $\pm$  0.04 &  0.81 $\pm$  0.02 &  406 & -0.35 \\
\hline
 56052.83425 & Secondary & 0.493 &  3.49 $\pm$  0.10 &  2.20 $\pm$  0.13 &  162 & -0.32 &  1.27 $\pm$  0.09 &  0.84 $\pm$  0.04 &  189 & -0.28 \\
 56052.83794 & Secondary & 0.496 &  3.36 $\pm$  0.11 &  2.26 $\pm$  0.12 &  171 & -0.33 &  1.33 $\pm$  0.08 &  0.84 $\pm$  0.04 &  195 & -0.23 \\
 56052.84162 & Secondary & 0.499 &  3.81 $\pm$  0.10 &  2.39 $\pm$  0.12 &  168 & -0.26 &  1.41 $\pm$  0.08 &  0.84 $\pm$  0.04 &  193 & -0.25 \\
 56052.84531 & Secondary & 0.501 &  3.53 $\pm$  0.12 &  2.34 $\pm$  0.11 &  184 & -0.30 &  1.23 $\pm$  0.07 &  0.85 $\pm$  0.03 &  210 & -0.27 \\
 56052.84992 & Secondary & 0.505 &  3.34 $\pm$  0.09 &  2.22 $\pm$  0.09 &  231 & -0.34 &  1.17 $\pm$  0.06 &  0.90 $\pm$  0.03 &  261 & -0.26 \\
 56052.86108 & Secondary & 0.514 &  3.40 $\pm$  0.09 &  2.40 $\pm$  0.12 &  194 & -0.32 &  0.96 $\pm$  0.06 &  0.85 $\pm$  0.03 &  225 & -0.32 \\
 56052.86477 & Secondary & 0.517 &  3.27 $\pm$  0.08 &  2.29 $\pm$  0.10 &  238 & -0.34 &  0.98 $\pm$  0.05 &  0.87 $\pm$  0.03 &  268 & -0.30 \\
\hline
 56054.10690 & Secondary & 0.496 &  3.51 $\pm$  0.10 &  2.24 $\pm$  0.12 &  172 & -0.31 &  1.04 $\pm$  0.08 &  0.89 $\pm$  0.04 &  197 & -0.30 \\
 56054.10920 & Secondary & 0.498 &  3.41 $\pm$  0.10 &  2.48 $\pm$  0.11 &  177 & -0.31 &  1.19 $\pm$  0.08 &  0.81 $\pm$  0.04 &  202 & -0.33 \\
 56054.11150 & Secondary & 0.500 &  3.55 $\pm$  0.11 &  2.11 $\pm$  0.13 &  172 & -0.32 &  1.21 $\pm$  0.08 &  0.83 $\pm$  0.05 &  197 & -0.27 \\
 56054.11381 & Secondary & 0.502 &  3.41 $\pm$  0.10 &  2.35 $\pm$  0.12 &  186 & -0.32 &  1.18 $\pm$  0.07 &  0.87 $\pm$  0.03 &  213 & -0.31 \\
 56054.11611 & Secondary & 0.503 &  3.42 $\pm$  0.33 &  2.37 $\pm$  0.24 &  193 & -0.31 &  1.18 $\pm$  0.11 &  0.82 $\pm$  0.17 &  217 & -0.32 \\
 56054.11842 & Secondary & 0.505 &  3.46 $\pm$  0.10 &  2.31 $\pm$  0.12 &  187 & -0.31 &  1.18 $\pm$  0.07 &  0.84 $\pm$  0.04 &  215 & -0.32 \\
 56054.12129 & Secondary & 0.507 &  3.35 $\pm$  0.08 &  2.28 $\pm$  0.11 &  211 & -0.33 &  1.10 $\pm$  0.07 &  0.91 $\pm$  0.03 &  240 & -0.31 \\
 56054.12389 & Secondary & 0.509 &  3.55 $\pm$  0.09 &  2.52 $\pm$  0.10 &  239 & -0.28 &  1.27 $\pm$  0.06 &  0.86 $\pm$  0.05 &  271 & -0.28 \\
\enddata
%\tablecomments{The list of metallicity measurements performed on CM Dra.}
\tablenotetext{a}{Average RJD of combined exposure.}
\tablenotetext{b}{Phase calculated using elements from \citet{mor2009}.}
\end{deluxetable*}

\begin{deluxetable}{cccccc}
\tabletypesize{\small}
\tablecaption{Averaged CM Dra Metallicity Measurements per Epoch\label{summarytable}}
\tablewidth{0pt}
\tablehead{
\colhead{Epoch} & \colhead{RJD} & \colhead{$\overline{\rm{[Fe/H]}}_{\mathrm{K}}$} & \colhead{$\sigma_{\rm{K}}$} & \colhead{$\overline{\rm{[Fe/H]}}_{\mathrm{H}}$} & \colhead{$\sigma_{\rm{H}}$}
}
\startdata
Out of eclipse 1 & 56055 & -0.30 & \ldots & -0.40 & \ldots \\
Out of eclipse 2 & 56056 & -0.33 & \ldots & -0.37 & \ldots \\
Primary Eclipse & 56050 & -0.33 &  0.02 & -0.36 &  0.02 \\
Secondary Eclipse 1 & 56052 & -0.32 &  0.03 & -0.27 &  0.03 \\
Secondary Eclipse 2 & 56054 & -0.31 &  0.01 & -0.31 &  0.02 \\
\enddata
\tablecomments{The summary by epoch of the metallicity measurements performed on CM Dra. The metallicity measurements are the means and standard deviations calculated for each epoch of observation.}
\end{deluxetable}

Interestingly, the primary eclipse and out-of-eclipse metallicity estimates do not deviate significantly from the measurement of the secondary eclipse. Since these stars are closely bound and are of similar mass (7\% difference) and spectral type (both dM4.5), this supports the validity of the assumption that the two stars are coeval, since differing compositions would likely yield different feature strengths between individual and combined light spectra.

The differences that do exist between the secondary eclipse observations and the primary/out-of-eclipse observations are only present in the $H$-band estimate, and are strongly correlated with airmass: when airmass is used to linearly model [Fe/H]$_{\rm{H-band}}$, the multiple correlation coefficient $R^{2}=0.78$. These differences are therefore likely byproducts of the sensitivity of the $H$-band technique to atmospheric contamination, as noted in \citet{ter2012}. In addition to the sensitivity of the \ion{K}{1} feature noted in the previous paper, in performing the CM Dra analysis we have also discovered a significant sensitivity of the \ion{Ca}{1} feature to a step in the {\it xtellcor} routine. Specifically, the user must manually scale the strengths of Brackett lines in a model spectrum of Vega in order to remove them from the observed telluric standard. It is not immediately obvious how to correctly scale a particular Brackett line in the vicinity of both the $H$-band \ion{Ca}{1} feature and a telluric absorption feature. Repeated iterations of our analysis allowed us to develop a consistent scaling based on requiring a certain smoothness at the edge of a telluric CO absorption feature across the multiple telluric standard observations for CM Dra. We are studying methods of dealing with the dependency of the $H$-band technique on telluric contamination using high-resolution $H$-band spectra from the SDSS-III APOGEE instrument \citep{all2008}.

\begin{figure}
\begin{center}
\includegraphics[scale=0.45]{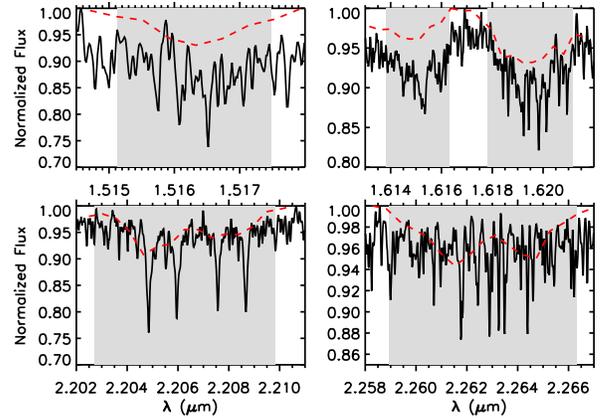}
\caption{Shifted spectral windows (from \citet{ter2012}) used to calculate the EWs (grey regions) compared to a pair of BT Settl models of $T_{\rm{eff}} = 3100 \: \rm{K}$, $\log{g} = 5.0$ and $\rm{[M/H]} = -0.5$, Doppler-shifted to the RVs expected at the time of the first out-of-eclipse observation and convolved to $R = 60,000$ (black).  The red, dashed line is the observed $R=2000$ IRTF spectrum for comparison.\label{specwindows}}
\end{center}
\end{figure}

Since CM Dra has both a high absolute RV ($\approx 100\rm{\ km\ s}^{-1}$) and high orbital RV amplitude ($\approx 70 \rm{\ km\ s}^{-1}$), it is important to ensure that the windows used to calculate feature strengths are shifted appropriately in RV (the window widths remain the same). Figure \ref{specwindows} shows the May 7 out-of-eclipse observations (near maximum RV separation), the feature windows shifted by our analysis routine, and an RV-shifted BT-Settl spectrum for reference \citep{all2003,all2007,all2011}.

Our estimate of the metallicity of CM Dra is the first in a long series of composition studies to converge on a precise answer. Comparison of individual features with models in both the optical and NIR \citep[e.g.\ atomic features and a CO band,][]{vit1997,vit2002} have concluded that CM Dra is metal poor ([M/H] $\approx -1.0$), which would be consistent with it being a halo/population II object. However, similar studies of other features \citep[e.g.\ CaH and TiO,][]{giz1997} and IR colors \citep{leg1998} have preferred a near-solar composition ([M/H] $\approx 0.0$), and multiple studies \citep{vit1997,leg1998} have noted difficulty in reconciling metallicity determinations for different wavelength regions. Recent studies have even included attempts at detailed line modeling \citep{kuz2012}. Our finding that CM Dra is slightly metal poor ([Fe/H] $\approx -0.30$) lies between the aforementioned studies that find a range of metallicities between metal poor and solar, and is not consistent with studies that find the system to be metal poor. Unlike these previous studies, our empirical estimate of the metallicity of CM Dra does not suffer from the difficulties of applying low-mass stellar atmospheric models, such as missing or poorly defined opacities. There is one published metallicity estimate using a similar technique to ours, from \citet{roj2012}, which finds [Fe/H] $= -0.39 \pm 0.17$ dex using features in the $K$-band. This estimate is consistent with ours, but to the best of our knowledge is not timed for a specific epoch and so likely contains mixed light. 

\section{Impact on Stellar Model Interpretation}
\label{modelsection}
\begin{figure}
\begin{center}
\includegraphics[scale=.48]{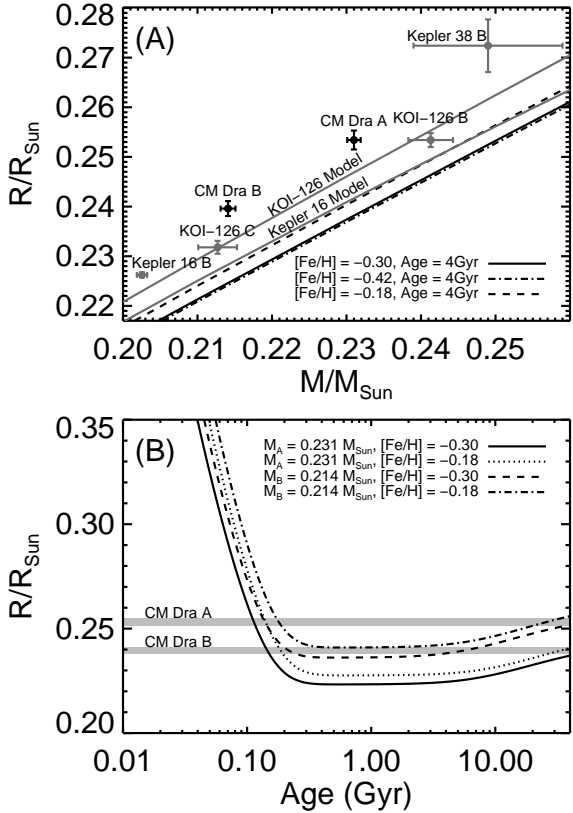}
\caption{Stellar models from the Dartmouth Stellar Evolution Program \citep{fei2012}. (A) Mass-radius diagram of stars that have $0.2 < M < 0.25 \, M_{\odot}$ and precise masses and radii. CM Dra A/B are shown in black, along with the DSEP isochrones for an age of 4 Gyr and our estimated metallicity values [Fe/H]$=-0.30 \pm 0.12$. Also plotted in gray are Kepler 16B and KOI-126B/C, along with their respective best-fit isochrones, from \citet{fei2012}. The recently-discovered Kepler 38 B \citep{oro2012}, which will be a strong test of isochrones when its stellar parameters are more tightly constrained, is also shown. (B) The evolutionary tracks for both components of CM Dra with the measured metallicity ([Fe/H] $=-0.3$), along with the same tracks for our 1-$\sigma$ upper limit metallicity estimate ([Fe/H] $=-0.18$). The tracks for the 1-$\sigma$ lower limit ([Fe/H] $=-0.42$) are indistinguishable from those for [Fe/H] $=-0.30$ in this plot.\label{isochrones}}
\end{center}
\end{figure}

Since CM Dra is among the lowest mass systems with well-determined stellar parameters, it is an important benchmark for low-mass stellar models, and will play an important role in work to reconcile the discrepancy between low-mass models and observations. The most recent study to address stellar models and CM Dra, \citet{fei2012}, still finds a discrepancy of $\sim 3\%$ between the observed and modeled radius for both components. This was derived using a metallicity prior of $-1 \leq \rm{[Fe/H]} \leq 0$, which allowed a best-fit metallicity of 0 as input to the Dartmouth Stellar Evolution Program \citep[DSEP,][]{dot2007,dot2008,fei2011,fei2012}. Figure \ref{isochrones} A shows a 4 Gyr DSEP isochrone with [Fe/H] = -0.3. The model radii are $R_{\rm{A,model}}=0.2374~R_{\rm{\odot}}$ and $R_{\rm{B,model}}=0.2213~R_{\rm{\odot}}$. Using the radii measured by \citet{mor2009}, the relative errors, defined as
\begin{equation}
\frac{\delta R}{R_{obs}} = \frac{R_{obs} - R_{model}}{R_{obs}},
\end{equation}
are $\delta R_{A}/R_{obs} = 0.063$ and $\delta R_{B}/R_{obs} = 0.076$. If we take instead the 1-$\sigma$ upper limit for metallicity of [Fe/H]$= -0.18$, we find that the discrepancies decrease to $\delta R_{A}/R_{obs} = 0.052$ and $\delta R_{B}/R_{obs} = 0.065$. Higher metallicities yield larger radii \citep[somewhat ameliorating the problem, as in e.g.][]{spa2012}, so the true extent of the discrepancy for CM Dra was almost certainly underestimated in \citet{fei2012} due to the relatively high allowed metallicity of 0.

Besides metallicity, another significant constraint on the stellar models is age. The age of the CM Dra system is supposedly well-constrained by the cooling age of its white dwarf companion \citep{ber2001,mor2009}. Figure \ref{isochrones} B shows the evolutionary tracks for models with the masses of CM Dra A/B and reveals the extent to which age can affect the radius discrepancy for this system. With [Fe/H] $=-0.30\pm0.12$, CM Dra B can only be realistically fit with ages of 100-200 Myr. However, the estimated 4 Gyr age of the white dwarf companion and the high space velocity of the system suggest that this system is much older. We conclude that with constraints on both age and metallicity, the non-magnetic models for this system remain discrepant at the 5-7\% level.

With the present constraint on metallicity and a long history of photometric and spectroscopic monitoring and modeling, CM Dra provides a strong test of low-mass stellar models. KOI-126 \citep{car2011,fei2011}, Kepler 16 \citep{doy2011,win2011} and  LSPM J1112+7626 \citep{irw2011} provide constraints in the fully-convective region as well, as will the recently-discovered Kepler 38 B \citep{oro2012}, when a larger number of observations are able to constrain its mass and radius to the $\sim 1$\% level. The discovery papers for both KOI-126 and Kepler 16 \citep{car2011, doy2011} provide spectroscopic estimates of metallicity of the higher-mass components of the systems, and for KOI-126 these are supplemented by a model-derived age constraint from \citet{fei2011}. Although KOI-126 is well-fit by the appropriate stellar models, CM Dra and Kepler 16 still deviate significantly (5-7\% and 3\%, respectively). Kepler 16 is of particular note due to its well-constrained mass which has been independently spectroscopically verified \citep{ben2012}.  LSPM J1112+7626B still deviates at the 3\% level as well, although its metallicity and age are not as well constrained \citep{irw2011,fei2012}. As observations of these objects tighten the constraints on stellar models, the discrepancy between models and observations persists, possibly pointing the way toward new physics that must be included in the models, as mentioned in Section \ref{introsection}. Importantly, CM Dra is one of only two known systems with well-constrained masses and radii (along with KOI-126) that are composed of low-mass stars in very short ($\approx$ 1 day) periods, and that will be able to provide direct tests of models which explain the observed inflated radii of low-mass stars with magnetic fields produced by rotationally-powered dynamos. A thorough understanding of low-mass stars, and especially well-constrained mass-radius relations, will also be particularly important for M dwarf planet searches \citep{nut2008,mah2010,qui2010}, as radius is a crucial input in the calculation of the location of the habitable zone \citep[e.g.][]{boy2012}. 
\section{Conclusion}
\label{conclusionsection}
We have estimated the metallicity of the CM Dra system using two empirical NIR relations applied to observations in and out of eclipse. We find it to be slightly metal poor, with [Fe/H]$= -0.30 \pm 0.12$ dex. This measurement is consistent with estimates obtained out of eclipse and during primary eclipse, suggesting that the components are very similar in composition, since they are close in mass and spectral type. When used as input to DSEP stellar models, this study and studies of the similar systems KOI-126, Kepler 16, and LSPM J1112+7626 show that the discrepancies between measured and predicted radiii persist at the 3-7\% level. The precise determination of its metallicity, as well as the constraint on its age from the proposed white dwarf companion, enables CM Dra to be one of the best test cases for low-mass stellar models. Further improvement in the understanding of low-mass stars will require both observational work in finding and characterizing low-mass stars, and theoretical work in modeling the additional physics necessary to match the observed cases.

\acknowledgements
We acknowledge support from NASA, NAI and PSARC, as well as the NSF through grants AST 1006676 and AST 1126413. This work was partially supported by funding from the Center for Exoplanets and Habitable Worlds. G.A.F. thanks the William H. Neukom 1964 Institute for Computational Science for their support. The Center for Exoplanets and Habitable Worlds is supported by the Pennsylvania State University, the Eberly College of Science, and the Pennsylvania Space Grant Consortium. Data obtained at the Infrared Telescope Facility, which is operated by the University of Hawaii under Cooperative Agreement no. NNX-08AE38A with the National Aeronautics and Space Administration, Science Mission Directorate, Planetary Astronomy Program. This research has made use of the SIMBAD database, operated at CDS, Strasbourg, France.

%%%%%%%%%%%%%%%%%% tables here %%%%%%%%%%%%%%%%%%

\clearpage

%%%%%%%%%%%%%%%%%% end figures %%%%%%%%%%%%%%%%%%


\begin{thebibliography}{}
\bibitem[Allard et al.\ (2007)]{all2007} Allard, F., Allard, N.~F., Homeier, D., et al.\ 2007, \aap, 474, L21 
\bibitem[Allard et al.\ (2003)]{all2003} Allard, F., Guillot, T., Ludwig, H.-G., et al.\ 2003, Brown Dwarfs, 211, 325 
\bibitem[Allard et al.\ (2011)]{all2011} Allard, F., Homeier, D., \& Freytag, B.\ 2011, Astronomical Society of the Pacific Conference Series, 448, 91
\bibitem[Allende Prieto et al.\ (2008)]{all2008} Allende Prieto, C., Majewski, S.~R., Schiavon, R. et al.\ 2008, Astronomische Nachrichten, 329, 1018
\bibitem[Bender et al.\ (2012)]{ben2012} Bender, C.~F., Mahadevan, S., Deshpande, R., et al.\ 2012, \apjl, 751, L31
\bibitem[Bergeron et al.\ (2001)]{ber2001} Bergeron, P., Leggett, S.~K., \& Ruiz, M.~T.\ 2001, \apjs, 133, 413
\bibitem[Boyajian et al.\ (2012)]{boy2012} Boyajian, T.~S., von Braun, K., van Belle, G., et al.\ 2012, \apj, 757, 112
\bibitem[Carter et al.\ (2011)]{car2011} Carter, J.~A., Fabrycky, D.~C., Ragozzine, D., et al.\ 2011 Science, 331, 562
\bibitem[Chabrier et al.\ (2007)]{cha2007} Chabrier, G., Gallardo, J., \& Baraffe, I.\ 2007, \aap, 472, L17
\bibitem[Cushing et al.\ (2004)]{cus2004} Cushing, M.~C., Vacca, W.~D., \& Rayner, J.~T.\ 2004, \pasp, 116, 362
\bibitem[Dotter et al.\ (2007)]{dot2007} Dotter, A., Chaboyer, B., Ferguson, J.~W. et al.\ 2007, \apj, 666, 403
\bibitem[Dotter et al.\ (2008)]{dot2008} Dotter, A., Chaboyer, B., Jevremovi\'{c}, D., et al.\ 2008, \apjs, 178, 89
\bibitem[Doyle et al.\ (2011)]{doy2011} Doyle, L.~R., Carter, J.~A., Fabrycky, D.~C., et al.\ 2011, Science, 333, 1602
\bibitem[Feiden et al.\ (2011)]{fei2011} Feiden, G.~A., Chaboyer, B., \& Dotter, A.\ 2011, \apjl, 740, L25
\bibitem[Feiden \& Chaboyer (2012)]{fei2012} Feiden, G.~A. \& Chaboyer, B.\ 2012, \apj, 757, 42
\bibitem[Gizis (1997)]{giz1997} Gizis, J.~E.\ 1997, \aj, 113, 806
\bibitem[Gliese \& Jahreiss (1991)]{gli1991} Gliese, W., \& Jahreiss, H. 1991, Astronomical Data Center CD-ROM: Selected Astronomical Catalogs, Vol. 1, ed. L.~E. Bratzmann \& S.~E. Gesser (Greenbelt, MD: NASA/GSFC)
\bibitem[Irwin et al.\ (2011)]{irw2011} Irwin, J.~M., Quinn, S.~N., Berta, Z.~K., et al.\ 2011, \apj, 742, 123
\bibitem[Kuznetsov et al.\ (2012)]{kuz2012} Kuznetsov, M.~K., Pavelnko, Ya.~V., Jones, H.~R.~A., Pinfield, D.~Dz.\ 2012, Advances in Astronomy and Space Physics, 2, 15
\bibitem[Lacy\ (1977)]{lac1977} Lacy, C.~H.\ 1977, \apj, 218, 444
\bibitem[Leggett (1998)]{leg1998} Legget, S.~K., Allard, F., Hauschildt, P.~H.\ 1998, \apj, 509, 836
\bibitem[L\'{o}pez-Morales (2007)]{lop2007} L\'{o}pez-Morales, M.\ 2007, \apj, 660, 732
\bibitem[MacDonald \& Mullan (2012)]{mac2012} MacDonald, J., \& Mullan, D.~J.\ 2012, \mnras, 421, 3084
\bibitem[Mahadevan et al.\ (2010)]{mah2010} Mahadevan, S., Ramsey, L.~W., Wright, J.~T., et al.\ 2010, Proc. SPIE, 7735
\bibitem[Morales et al.\ (2008)]{mor2008} Morales, J.~C., Ribas, I., \& Jordi, C.\ 2008, \aap, 478, 507
\bibitem[Morales et al.\ (2009)]{mor2009} Morales, J.~C., Ribas, I., Jordi, C., et al.\ 2009, \apj, 691, 1400
\bibitem[Morales et al.\ (2010)]{mor2010} Morales, J.~C., Gallardo, J., Ribas, I., et al.\ 2010, \apj, 718, 502
\bibitem[Nutzman \& Charbonneau (2008)]{nut2008} Nutzman, P., \& Charbonneau, D.\ 2008, \pasp, 120, 317
\bibitem[Orosz et al.\ (2012)]{oro2012} Orosz, J.~A., Welsh, W.~F., Carter, J.~A. et al.\ 2012, \apj, in press, arXiv:1208.3712
\bibitem[Quirrenbach et al.\ (2010)]{qui2010} Quirrenbach, A., Amado, P.~J., Mandel, H., et al.\ 2010, Proc. SPIE, 7735
\bibitem[Rayner et al.\ (2003)]{ray2003} Rayner, J.~T., Toomey, D.~W., Onaka, P.~M., et al.\ 2003, \pasp, 115, 3
62
\bibitem[Rayner et al.\ (2009)]{ray2009} Rayner, J.~T., Cushing, M.~C., Vacca, W.~D. 2009, \apjs, 185, 289
\bibitem[Reiners et al.\ (2012)]{rei2012} Reiners, A., Joshi, N., Goldman, B. 2012, \apj, 143, 93
\bibitem[Ribas (2006)]{rib2006} Ribas, I.\ 2006, \apss, 304, 89
\bibitem[Rojas-Ayala et al.\ (2010)]{roj2010} Rojas-Ayala, B., Covey, K.~R., Muirhead, P.~S., \& Lloyd, J.~P.\ 2010, \apjl, 720, L113
\bibitem[Rojas-Ayala et al.\ (2012)]{roj2012} Rojas-Ayala, B., Covey, K.~R., Muirhead, P.~S., \& Lloyd, J.~P.\ 2012, \apj, 748, 93
\bibitem[Spada \& Demarque (2012)]{spa2012} Spada, F., \& Demarque, P.\ 2012, \mnras, 422, 2255
\bibitem[Stassun et al.\ (2012)]{sta2012} Stassun, K.~G., Kratter, K.~M., Scholz, A., Dupuy, T.~J.\ 2012, \apj, 756, 47
\bibitem[Terrien et al.\ (2012)]{ter2012} Terrien, R.~C., Mahadevan, S., Bender, C.~F., Deshpande, R., Ramsey, L.~W., Bochanski, J.~J.\ 2012, \apjl, 747, L38
\bibitem[Torres et al.\ (2010)]{tor2010} Torres, G., Andersen, J., \& Gim{\'e}nez, A.\ 2010, \aapr, 18, 67
\bibitem[Vacca et al.\ (2003)]{vac2003} Vacca, W.~D., Cushing, M.~C., \& Rayner, J.~T.
\bibitem[Valenti \& Fischer (2005)]{val2005} Valenti, J.~A., Fischer, D.~A. 2005, \apjs, 159, 141
\bibitem[Viti et al.\ (1997)]{vit1997} Viti, S., Jones, H.~R.~A., Schweitzer, et al.\ 1997, \mnras, 291, 780
\bibitem[Viti et al.\ (2002)]{vit2002} Viti, S., Jones, H.~R.~A., Maxted, P., \& Tennyson, J.\ 2002, \mnras, 329, 290
\bibitem[Winn et al.\ (2011)]{win2011} Winn, J.~N., Albrecht, S., Johnson, J.~A., et al.\ 2011, \apjl, 741, L1
\bibitem[Wright \& Howard (2009)]{wri2009} Wright, J.~T., \& Howard, A.~W. 2009, \apjs, 182, 205
\end{thebibliography}
\end{document}